\documentclass[9pt,twocolumn,twoside]{osajnl}
\usepackage{tabu}

\newcommand{\ind}{\hspace{0.2 in}}

\journal{ol} 

\setboolean{shortarticle}{true}
\usepackage{verbatim}  
\usepackage{ulem} 
\usepackage{float}
\usepackage[export]{adjustbox}

\title{Mid-infrared frequency comb with 6.7 W average power based on difference frequency generation}

\author[1,*]{Anthony Catanese}
\author[1,*]{Jay Rutledge}
\author[1]{Myles Silfies}
\author[1]{Xinlong Li}
\author[2]{Henry Timmers}
\author[2]{Abijith S. Kowligy}
\author[2] {Alex Lind}
\author[2,3]{Scott A. Diddams}
\author[1,+]{Thomas K. Allison}

\affil[1]{Stony Brook University, Stony Brook, NY 11790, USA}
\affil[2]{National Institute of Standards and Technology, Boulder, CO 80305, USA}
\affil[3]{Department of Physics, University of Colorado Boulder, Boulder, CO 80309, USA}
\affil[+]{thomas.allison@stonybrook.edu}
\affil[*]{Both authors contributed equally to this work.}
    
\begin{abstract}

We report on the development of a high-power mid-infrared frequency comb with 100 MHz repetition rate and 100 fs pulse duration. Difference frequency generation is realized between two branches derived from an Er:fiber comb, amplified separately in Yb:fiber and Er:fiber amplifiers. Average powers of 6.7 W and 14.9 W are generated in the 2.9 $\mu$m idler and 1.6 $\mu$m signal, respectively. With high average power, excellent beam quality, and passive carrier-envelope phase stabilization, this light source is a promising platform for generating broadband frequency combs in the far infrared, visible, and deep ultraviolet. 

\end{abstract}

\setboolean{displaycopyright}{true}

\begin{document}

\maketitle


The extension of optical frequency comb techniques to the mid-infrared (mid-IR) has been mainly motivated by applications in molecular spectroscopy, including spectroscopy in the "fingerprint" region \cite{Changala_Science2019, Timmers_Optica2018} and trace-gas sensing with high discrimination \cite{Foltynowicz_APB2013, NugentGlandorf_ApplPhysB2015}. The development of mid-infrared combs at the $<$1 W level for molecular spectroscopy has thus been an area of intense activity. A variety of mid-IR frequency comb technologies have emerged over the last decade, among them mode-locked quantum cascade lasers \cite{Hugi_Nature2012}, microresonator combs \cite{Luke_OptLett2015}, and combs based on  optical parametric amplifiers (OPA) \cite{Ruehl_OptLett2012} and optical parametric oscillators (OPO) \cite{Adler_OptLett2009}. 

\ind In addition to spectroscopy, the mid-infrared is also an excellent region for driving nonlinear optics. The 2-5 $\mu$m wavelength range is particularly attractive because it offers advantages for generating both coherent ultraviolet/visible light via high harmonic generation (HHG) \cite{Hickstein_Optica2017, Popmintchev_Science2012, Ghimire_NatPhys2019}, and broadband infrared combs via difference frequency generation \cite{Vasilyev_Optica2019, Gaida_LightSciAdv2018} or supercontinuum generation \cite{Seidel_SciAdv2018}. For HHG in gases, the mid-IR offers extension of the cutoff photon energy via scaling of the pondermotive energy \cite{Popmintchev_Science2012}. For HHG in solids, the low mid-IR photon energy enables driving nonperturbative photocurrents \cite{Ghimire_NatPhys2019} or cascaded $\chi^{(2)}$ and $\chi^{(3)}$ processes \cite{Hickstein_Optica2017} without damage due to multiphoton absorption. For broadband infrared continuum generation via intra-pulse DFG, using driving wavelengths in the mid-infrared enjoys much better group velocity matching compared to driving with more conventional near-IR combs, enabling high average powers and super-octave coherent infrared bandwidths \cite{Gaida_LightSciAdv2018, Vasilyev_Optica2019}. In these ways, the 2-5 $\mu$m, spectral range offers a pivot to generating both shorter and longer wavelengths.

\ind For driving nonlinear optics efficiently at high repetition rate, high average powers are needed. For this application, frequency combs based on DFG then offer several advantages. For power scaling, no heat (ideally) is deposited in parametric gain media and with DFG combs based on multiple branches, the separate branchs can be amplified to high average powers in fiber amplifiers \cite{Cruz_OptExp2015, Maser_ApplPhysB2017, Zhu_OptLett2013}. For nonperturbative nonlinear optics sensitivite to the electric field (e.g. solid-state HHG), OPAs based on difference frequency generation (DFG) between a pump and signal derived from the same comb are particularly attractive due to passive elimination of the carrier-envelope offset frequency, $f_0$, producing a train of carrier-envelope phase (CEP) stable pulses. This stands in contrast to competing technologies based on Cr:ZnS \cite{Vampa_OptLett2019} or Tm:fiber \cite{Lee_OptLett2017}, for which it is nontrivial to stabilize $f_0$ to zero.

\begin{figure*}[ht]
\centering
\fbox{\includegraphics[width=0.98\linewidth]{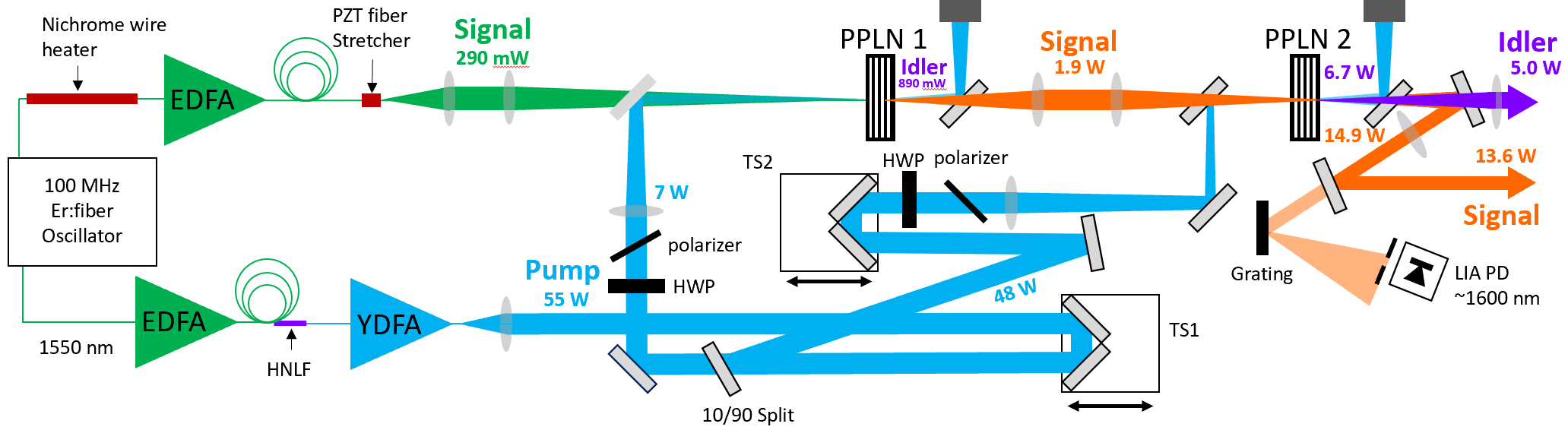}}
\caption{Schematic of the two stage OPA. Pump and signal branches are derived from an Er:fiber oscillator are amplified in separate Er:fiber (EDFA) and Yb:fiber (YDFA) amplifiers. Difference frequency generation in two periodically-poled lithium niobate crystals (PPLN1 and PPLN2) generates the high power signal and idler combs. Path-length stabilization is achieved by heating the fiber in the signal branch with a nichrome heater wire. Translation stage TS1 controls the pump/signal delay in both OPA stages, with TS2 changing only stage 2. More details in the text.}
\label{fig:OPA_Design}
\end{figure*}

\ind In this letter, we present  a 2.9 $\mu$m frequency comb based on high-power DFG in a two-stage OPA, operating at 100 MHz repetition rate with 6.7 W average power and 100 fs pulse duration. To our knowledge, this is the highest power mid-IR optical frequency comb reported to date and among the highest average power ultrafast mid-IR light sources in general. Furthermore, we demonstrate that this OPA can generate sufficient bandwidth to support few-cycle mid-IR pulses with further refinement of the seeding signal branch.

\ind A schematic of our design is shown in figure \ref{fig:OPA_Design}. The pump (1035 nm) and signal (1500-1650 nm) are both derived from an all polarization-maintaining (PM) fiber Er:fiber oscillator (Menlo Systems M-comb ultralow noise variant) with a center wavelength of 1560 nm. In addition to convenient pulse synchronization, deriving both pump and signal combs from the same oscillator ensures the idler comb generated via DFG has $f_0 = 0$.

\ind The pump branch uses an intricate yet robust chain of nonlinear fiber optics to shift the Er:fiber comb from 1560 nm to 1035 nm \cite{Ycas_OptLett2012}. From the Er:fiber oscillator, the light is bandpass-filtered (10 nm) and amplified in a nonlinear Erbium-doped fiber amplifier (EDFA) incorporating normal-dispersion Er-doped fiber (1.5 m Er80-4/125-HD-PM, D $\approx$ -22 ps/nm/km, 80 dB/m absorption at 1530 nm)  pumped by four 750 mW, 976 nm pump diodes. After a length of the anomalous dispersion fiber after the EDFA (30 cm, D $\approx$ 18 ps/nm/km), sub-50 fs pulses with 350 mW average power enter a 3 cm long piece of anomalous dispersion (D = 5.6 ps/nm/km, $\gamma =$ 10.5 W$^{-1}$km$^{-1}$) highly nonlinear fiber directly spliced to the anomalous dispersion fiber. Dispersive wave generation in the HNLF gives a comb with $\sim$15 mW of power between 1000 and 1100 nm. All components of this fiber assembly are polarization maintaining (PM) for excellent long-term stability. The output of the HNLF is subsequently amplified to 200 mW in a nonlinear, single-mode fiber, core-pumped, Yb-doped fiber amplifier (YDFA) and stretched to $\sim$100 ps using an anomalous third-order dispersion fiber stretcher \cite{Li_RSI2016}. This light is then used to seed a two stage high-power chirped-pulse YDFA system previously described in \cite{Li_RSI2016}, giving up to 55 W and 180 fs pulses after grating-pair compression.

\begin{figure*}[ht]
\centering
\fbox{\includegraphics[width=0.98\textwidth]{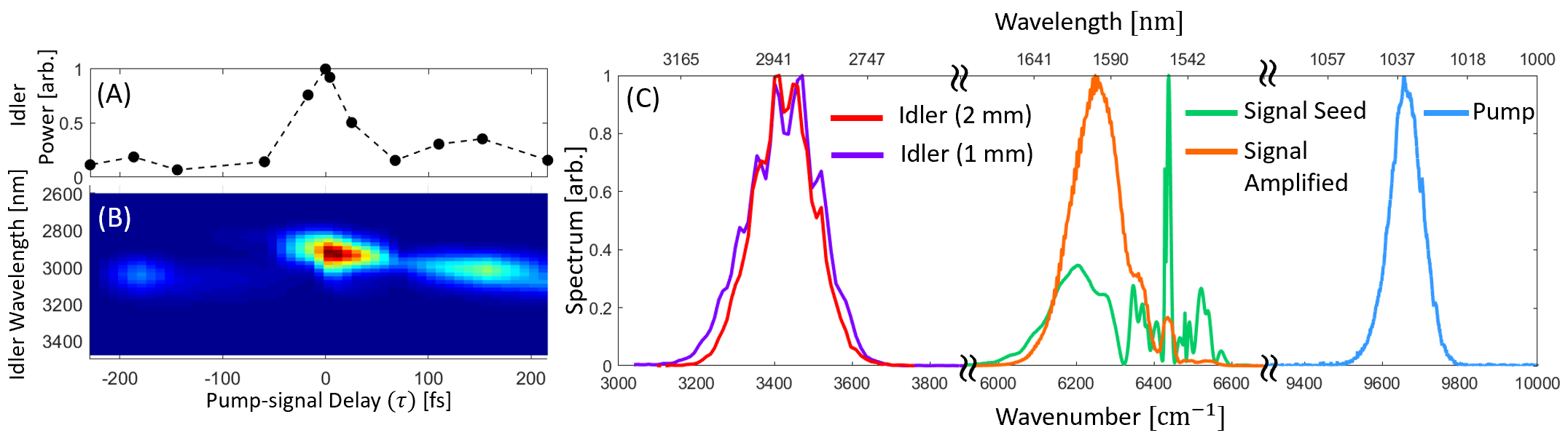}}
\caption{a) and b) Normalized stage 2 idler power and idler spectrum vs. pump-signal delay $\tau$. Positive $\tau$ indicate that the pump is arriving after the signal. c) Spectrum of the pump, signal seed, signal output, and idler output of the OPA at $\tau = 0$. The two idler spectra shown are for 1 and 2 mm PPLN crystals in stage 2.}
\label{fig:Spectrum}
\end{figure*}

\ind The signal branch is comparatively much simpler. A similar nonlinear EDFA is used, but now with a longer 75 cm single-mode anomalous-dispersion fiber spliced to its output. This provides 290 mW and the complicated spectrum shown as the green curve of figure \ref{fig:Spectrum}c) for seeding the OPA. The total optical path length difference of $\sim$20 m between the two branches, with high-power fiber amplifiers in both arms of the interferometer, is subject to long-term thermal drift. To stabilize the delay between the pump and signal at the OPA, the signal branch is equipped with two actuators. A piezoelectric fiber stretcher, driven with a 1071 Hz 20 V amplitude sine wave, modulates the signal seed delay by approximately 1 fs, which results in 0.1\% amplitude modulation on the output of the OPA. This amplitude modulation is detected with a photodiode and a lock-in amplifier to give an error signal. The error signal is integrated using a micro-controller which controls heating current sent to 1 m of nichrome wire kapton-taped directly to the EDFA gain fiber. A current of $\sim$1 A increases the fiber temperature by approximately 30$^{\circ}$ C, and gives 1.1 ps of delay. With this setup, the path length can be stabilized indefinitely after an initial warm-up period.

\ind The OPA itself is done in two stages using 5\%-MgO-doped periodically-poled lithium niobate (PPLN). A 2 mm long PPLN crystal, with a poling period of 30.49 $\mu$m, was ultimately chosen for both stages, but we report results with both 1 mm and a 2 mm long crystals for PPLN2. The crystals are heated to 80$^{\circ}$ C to optimize phase matching and avoid photorefractive damage. Two stages offers several advantages for the high-power OPA. First, as discussed by Arisholm et al. \cite{Arisholm_JOSAB2004}, it is generally easier to achieve good beam quality in high-gain OPAs with multiple stages. Second, independent control of the pump/signal delay in each stage (via the translation stages shown in figure \ref{fig:OPA_Design}) allows partial compensation of the pump/signal group velocity walk-off in stage 1. The pump power of each OPA stage is independently adjustable using a combination of a half wave plate and thin film polarizer. With 55 W from the YDFA, up to 7 W and 48 W can be used to pump stages 1 and 2 respectively.  The pump light is focused to spot sizes of 78 $\mu$m and 161 $\mu$m (FWHM) in stage 1 and 2, respectively. The signal beam in each stage is focused to the same size as the pump. We have not observed crystal damage over several months of operation.

\ind Figures \ref{fig:Spectrum}a) and \ref{fig:Spectrum}b) show the OPA output power and idler spectrum as the pump/signal delay, $\tau$, for both OPA stages is varied using translation stage TS1, showing multiple pulses emerging from the anomalous dispersion fiber. $\tau = 0$ is taken to be the delay of highest idler power. Mid-IR spectra were acquired using a scanning Czerny-Turner monochromator and liquid-nitrogen cooled InSb photodetector. The highest output power is observed when the OPA output spectrum is centered at 2900 nm, with the corresponding amplified signal spectrum (orange curve of figure \ref{fig:Spectrum}c) centered at 1600 nm. This well isolated pulse at $\tau = 0$, trailing the main 1550 nm pulse at $\tau = -200$ fs corresponds to a Raman-shifted soliton generated in the long anomalous dispersion fiber pigtail of the signal-branch EDFA. Thus, despite the complicated temporal structure emerging from the simple signal branch fiber assembly, clean soliton pulses can be isolated for amplification in the OPA. For the rest of the paper, we present results recorded at $\tau = 0$.


\begin{figure}[ht]
  \centering
  \includegraphics[width=0.44\textwidth]{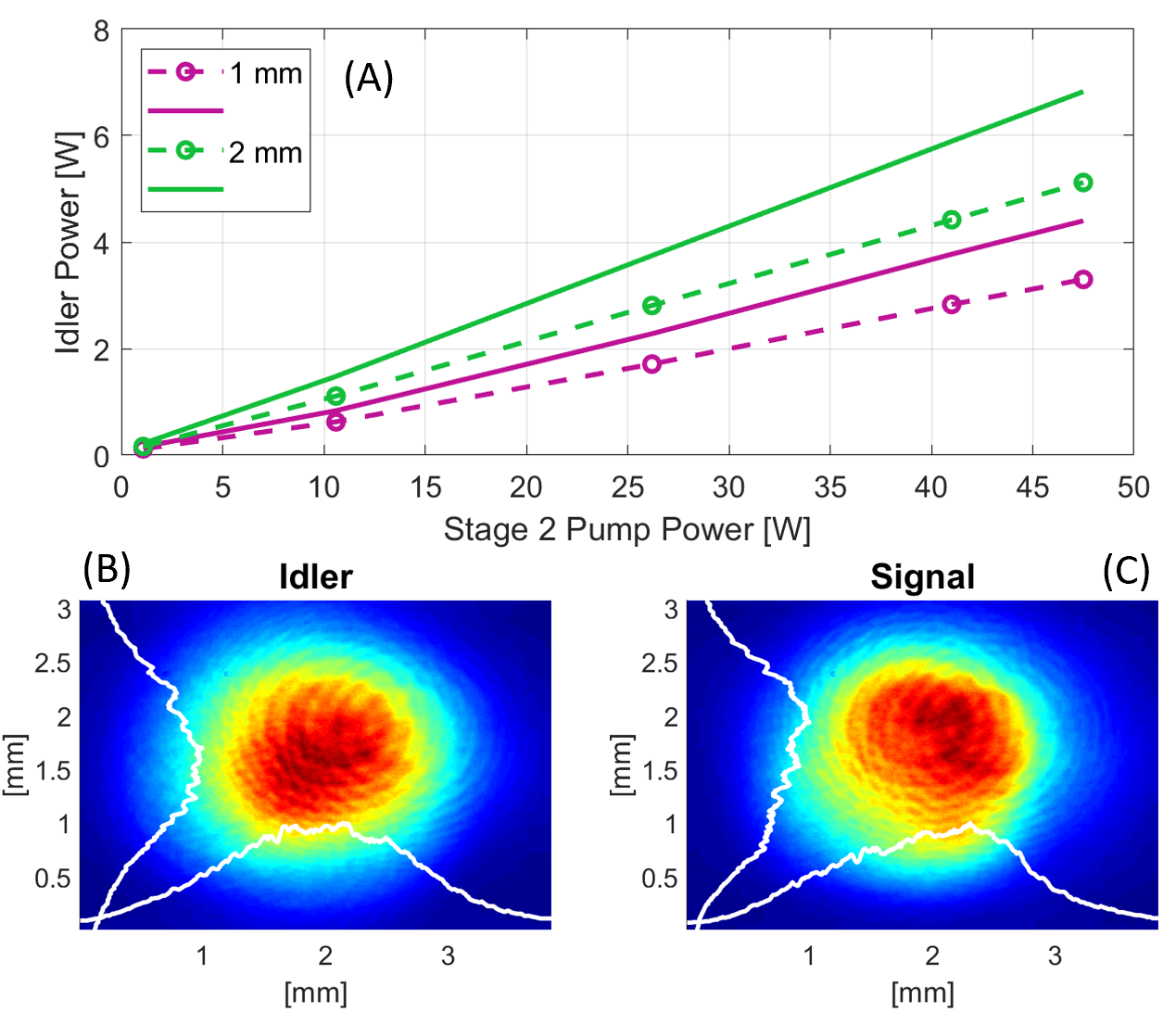}
 \caption{a) Stage 2 idler power vs. pump power for both 1 mm and 2 mm long crystals.  Dashed lines indicate the measured power after separating and collimating optics. b) Spatial mode profiles of the idler and the signal at full power.  Vertical and horizontal profiles along lineouts intersecting the centroid are shown in white.} 
\label{fig:powerscaling}
\end{figure}
 
\ind  When pumping the first stage with 7 W, more than 2 W of signal light and 900 mW of idler light emerge from PPLN1. Only the signal is retained between stage 1 and stage 2. With 48 W pump power in stage 2, at the exit of PPLN2 there is 6.7 W of idler and 14.9 W of signal. Figure \ref{fig:powerscaling}a) shows the idler output power vs. stage 2 pump power. Curves are shown for both 1 mm and 2 mm long PPLN crystals. The idler is isolated from the signal and pump via two dichroic mirrors and collimated with an f=25 cm CaF$_2$ lens. The dashed lines are the measured power after these output optics and represent the usable idler power from the OPA, while the the solid curves represent the output power corrected for the measured 25\% losses in the output optics. With the current output optics, the 2 mm (1 mm) long PPLN yields 5 W (3.3 W) usable idler power after the dichroic mirrors. Figure \ref{fig:powerscaling}b) also shows the output spot profiles at high power, measured using an additional 0.5x telescope and a microbolometer array camera. Despite the high powers involved, excellent beam quality is observed in both the signal and idler spatial modes. 
 
\begin{figure}[ht]
  \centering
  \includegraphics[width=0.45\textwidth]{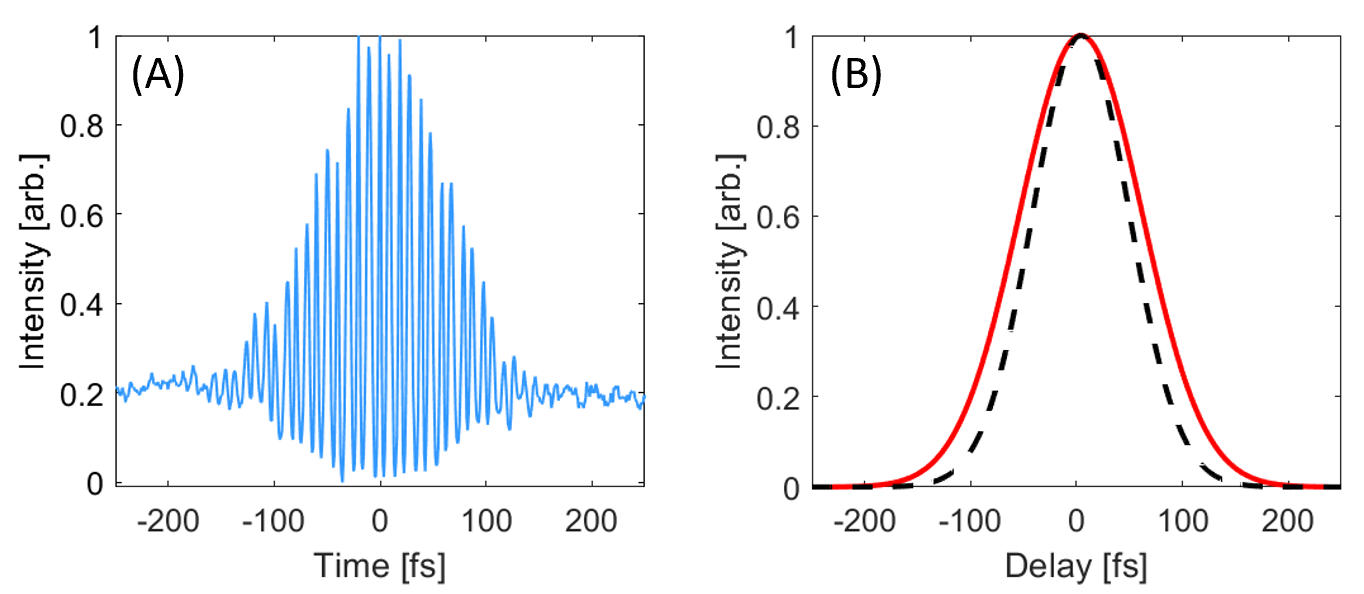}
 \caption{a) Idler intereferometric autocorrelation for the PPLN2 = 2 mm thick crystal. b) Intensity autocorrelation from low-pass filtering (red), which assuming a Gaussian pulse shape, corresponds to a measured idler pulse duration of 96 fs.  Calculated autocorrelation (black) for a pulse with the spectrum shown in figure \ref{fig:Spectrum}c).} 
\label{fig:AC}
 \end{figure}
  
\ind Figure \ref{fig:AC}a) shows an interferometric autocorrelation of the idler pulse, measured using a Michelson interferometer with a 2-photon InGaAs photodiode detector. Figure \ref{fig:AC}b) shows the intensity autocorrelation obtained from low-pass filtering the data. Assuming a Gaussian pulse shape, the measured idler pulse duration is determined to be 96 fs (FHWM). The transform-limited pulse calculated from the idler spectrum of figure \ref{fig:Spectrum}c) has a FWHM of 78 fs. For comparison with the experimental data, we also calculated the autocorrelation that the transform-limited idler pulses would have produced, and this is shown in dashed black. 
 
\begin{figure}[ht]
  \centering
  \includegraphics[width=0.48\textwidth,left]{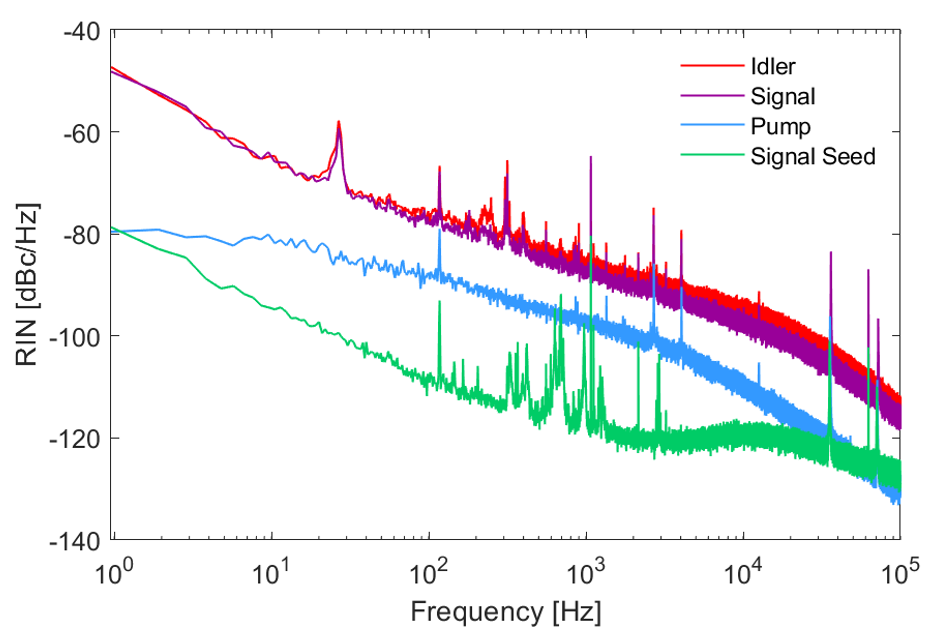}
 \caption{RIN for idler, signal output, pump, and signal seed.} 
\label{fig:RIN}
\end{figure}

\ind Figure \ref{fig:RIN} shows the relative intensity noise (RIN) in units of dBc/Hz for the idler output, signal output, pump, and signal seed. Analysis of the corresponding time series data indicated that idler and signal noise are strongly correlated, but uncorrelated from the noise of the pump and signal seed. Interestingly, we observed the RIN levels to not depend strongly on the pump/signal delay, in contrast to the recent reports of de Oliveira et al. \cite{deOliviera_arXiv2019}.

\ind Finally, we address the current OPA bandwidth and what is attainable with this platform. Figure \ref{fig:Spectrum}c) shows the idler spectrum for either the 1 and 2 mm PPLN crystals in the second OPA stage. The bandwidth obtained is nearly the same. Similarly, the pulse duration measured with 1 and 2 mm crystals is also the same within experimental error. This indicates that the bandwidth and pulse duration are not limited by the the group velocity walk-off in PPLN2, but instead by the fragmented signal seed pulse. Further evidence that the OPA can support much larger bandwidths is seen in figure \ref{fig:Spectrum}b), where output idler spectra are recorded as the pump/signal delay is varied. Parametric gain is achieved over $\sim$ 400 nm (450 cm$^{-1}$) of bandwidth with a single poling period and crystal temperature. To further explore this, we modeled the OPA in one spatial dimension as a $\chi^{(2)}$ three-wave mixing of temporally-Gaussian pulses according to \cite{Maser_ApplPhysB2017}. We applied the Fourier split-step method to solve two sequentially and independently pumped stages PPLN, with 25 fs transform-limited input seed pulses and 2 mm PPLN crystals in both stages. The results of the simulation predict that $>$400 nm phase-matched bandwidths supporting transform-limited pulses of less than 35 fs duration, or less than 3.5 optical cycles, can be attained. 

\ind In this letter we have presented the design and performance of a high-power mid-IR frequency comb based on difference frequency generation. The system has demonstrated the ability to produce 6.7 W of 2.9 $\mu$m idler light and 14.9 W of 1.6 $\mu$m signal light with excellent beam quality and measured idler pulse duration of 100 fs. We have also shown that the OPA is capable of supporting larger bandwidths and shorter, even few-cycle pulses, with further refinement of the seed branch. Very short 1550 nm pulses for seeding the OPA can be generated using a similar nonlinear EDFA seed branch by using short normal dispersion HNLFs, as in \cite{Timmers_Optica2018}, and we plan to implement this in the future. With few cycle-pulses and intrinsic CEP stability, this source is attractive for driving nonperturbative HHG in solids, with the possibility of generating isolated attosecond VUV pulses at high repetition rate via gating schemes \cite{Ghimire_NatPhys2019}. The high-power signal beam offers additional opportunities. Even with the current usable idler parameters of 5 W of average power and 100 fs pulse duration, focusing to 6 $\mu$m $(2\lambda)$ FWHM would yield peak intensities of $1.6\times10^{12}$ W/cm$^2$, sufficient to reach the nonperturbative HHG regime in a variety of crystals \cite{Ghimire_NatPhys2019}. We note that solid-state HHG at $\sim$100 MHz rate has recently been achieved using alternative high-power, but not CEP-stable, mid-IR sources \cite{Vampa_OptLett2019, Lee_OptLett2017}. Also, using this high power DFG comb, we have recently realized cascaded HHG in PPLN waveguides at 100 MHz for the generation of broadband visible frequency combs \cite{Kowligy-inPrep}. This OPA platform can also be used for tunable high-power mid-IR comb generation if dispersive-wave generation is implemented in the signal branch and the PPLN poling period tuned as in \cite{Maser_ApplPhysB2017}.

\vspace{0.2 cm}


\large{\textbf{Funding.}} National Science Foundation (NSF) (1708743); Air
Force Office of Scientific Research (AFOSR) (FA9550-16-1-0016 and FA9550-16-1-0164). Defense Advanced Research Projects Agency (DARPA) SCOUT program. M.C. Silfies and A. Catanese acknowledge support from the GAANN program of the U.S. Dept. of Education.

\bibliography{main}

\begin{thebibliography}{10}
\newcommand{\enquote}[1]{``#1''}

\bibitem{Changala_Science2019}
P.~B. Changala, M.~L. Weichman, K.~F. Lee, M.~E. Fermann, and J.~Ye,
  \enquote{Rovibrational quantum state resolution of the c60 fullerene,}
  {\protect\JournalTitle{Science}} \textbf{363}, 49--54 (2019).

\bibitem{Timmers_Optica2018}
H.~Timmers, A.~Kowligy, A.~Lind, F.~C. Cruz, N.~Nader, M.~Silfies, G.~Ycas,
  T.~K. Allison, P.~G. Schunemann, S.~B. Papp, and S.~A. Diddams,
  \enquote{Molecular fingerprinting with bright, broadband infrared frequency
  combs,} {\protect\JournalTitle{Optica}} \textbf{5}, 727--732 (2018).

\bibitem{Foltynowicz_APB2013}
A.~Foltynowicz, P.~Mas{\l}owski, A.~J. Fleisher, B.~J. Bjork, and J.~Ye,
  \enquote{Cavity-enhanced optical frequency comb spectroscopy in the
  mid-infrared application to trace detection of hydrogen peroxide,}
  {\protect\JournalTitle{Applied Physics B}} \textbf{110}, 163--175 (2013).

\bibitem{NugentGlandorf_ApplPhysB2015}
L.~Nugent-Glandorf, F.~R. Giorgetta, and S.~A. Diddams, \enquote{Open-air,
  broad-bandwidth trace gas sensing with a mid-infrared optical frequency
  comb,} {\protect\JournalTitle{Applied Physics B}} \textbf{119}, 327--338
  (2015).

\bibitem{Hugi_Nature2012}
A.~Hugi, G.~Villares, S.~Blaser, H.~C. Liu, and J.~Faist, \enquote{Mid-infrared
  frequency comb based on a quantum cascade laser,}
  {\protect\JournalTitle{Nature}} \textbf{492}, 229--233 (2012).

\bibitem{Luke_OptLett2015}
K.~Luke, Y.~Okawachi, M.~R.~E. Lamont, A.~L. Gaeta, and M.~Lipson,
  \enquote{Broadband mid-infrared frequency comb generation in a si3n4
  microresonator,} {\protect\JournalTitle{Optics Letters}} \textbf{40},
  4823--4826 (2015).

\bibitem{Ruehl_OptLett2012}
A.~Ruehl, A.~Gambetta, I.~Hartl, M.~E. Fermann, K.~S.~E. Eikema, and
  M.~Marangoni, \enquote{Widely-tunable mid-infrared frequency comb source
  based on difference frequency generation,} {\protect\JournalTitle{Opt.
  Lett.}} \textbf{37}, 2232--2234 (2012).

\bibitem{Adler_OptLett2009}
F.~Adler, K.~C. Cossel, M.~J. Thorpe, I.~Hartl, M.~E. Fermann, and J.~Ye,
  \enquote{Phase-stabilized, 1.5 w frequency comb at 2.8--4.8 $\mu$m,}
  {\protect\JournalTitle{Opt. Lett.}} \textbf{34}, 1330--1332 (2009).

\bibitem{Hickstein_Optica2017}
D.~D. Hickstein, D.~R. Carlson, A.~Kowligy, M.~Kirchner, S.~R. Domingue,
  N.~Nader, H.~Timmers, A.~Lind, G.~G. Ycas, M.~M. Murnane, H.~C. Kapteyn,
  S.~B. Papp, and S.~A. Diddams, \enquote{High-harmonic generation in
  periodically poled waveguides,} {\protect\JournalTitle{Optica}} \textbf{4},
  1538--1544 (2017).

\bibitem{Popmintchev_Science2012}
T.~Popmintchev, M.-C. Chen, D.~Popmintchev, P.~Arpin, S.~Brown, S.~Ali{\v
  s}auskas, G.~Andriukaitis, T.~Bal{\v c}iunas, O.~D. M{\"u}cke, A.~Pugzlys,
  A.~Baltu{\v s}ka, B.~Shim, S.~E. Schrauth, A.~Gaeta,
  C.~Hern{\'a}ndez-Garc{\'\i}a, L.~Plaja, A.~Becker, A.~Jaron-Becker, M.~M.
  Murnane, and H.~C. Kapteyn, \enquote{Bright coherent ultrahigh harmonics in
  the kev x-ray regime from mid-infrared femtosecond lasers,}
  {\protect\JournalTitle{Science}} \textbf{336}, 1287--1291 (2012).

\bibitem{Ghimire_NatPhys2019}
S.~Ghimire and D.~A. Reis, \enquote{High-harmonic generation from solids,}
  {\protect\JournalTitle{Nature Physics}} \textbf{15}, 10--16 (2019).

\bibitem{Vasilyev_Optica2019}
S.~Vasilyev, I.~S. Moskalev, V.~O. Smolski, J.~M. Peppers, M.~Mirov, A.~V.
  Muraviev, K.~Zawilski, P.~G. Schunemann, S.~B. Mirov, K.~L. Vodopyanov, and
  V.~P. Gapontsev, \enquote{Super-octave longwave mid-infrared coherent
  transients produced by optical rectification of few-cycle 2.5-$\mu$m pulses,}
  {\protect\JournalTitle{Optica}} \textbf{6}, 111--114 (2019).

\bibitem{Gaida_LightSciAdv2018}
C.~Gaida, M.~Gebhardt, T.~Heuermann, F.~Stutzki, C.~Jauregui, J.~Antonio-Lopez,
  A.~Sch{\"u}lzgen, R.~Amezcua-Correa, A.~T{\"u}nnermann, I.~Pupeza, and
  J.~Limpert, \enquote{Watt-scale super-octave mid-infrared intrapulse
  difference frequency generation,} {\protect\JournalTitle{Light: Science \&
  Applications}} \textbf{7}, 94 (2018).

\bibitem{Seidel_SciAdv2018}
M.~Seidel, X.~Xiao, S.~A. Hussain, G.~Arisholm, A.~Hartung, K.~T. Zawilski,
  P.~G. Schunemann, F.~Habel, M.~Trubetskov, V.~Pervak, O.~Pronin, and
  F.~Krausz, \enquote{Multi-watt, multi-octave, mid-infrared femtosecond
  source,} {\protect\JournalTitle{Science Advances}} \textbf{4} (2018).

\bibitem{Cruz_OptExp2015}
F.~C. Cruz, D.~L. Maser, T.~Johnson, G.~Ycas, A.~Klose, F.~R. Giorgetta,
  I.~Coddington, and S.~A. Diddams, \enquote{Mid-infrared optical frequency
  combs based on difference frequency generation for molecular spectroscopy,}
  {\protect\JournalTitle{Opt. Express}} \textbf{23}, 26814--26824 (2015).

\bibitem{Maser_ApplPhysB2017}
D.~L. Maser, G.~Ycas, W.~I. Depetri, F.~C. Cruz, and S.~A. Diddams,
  \enquote{Coherent frequency combs for spectroscopy across the 3--5 micron
  region,} {\protect\JournalTitle{Applied Physics B}} \textbf{123}, 142 (2017).

\bibitem{Zhu_OptLett2013}
F.~Zhu, H.~Hundertmark, A.~A. Kolomenskii, J.~Strohaber, R.~Holzwarth, and
  H.~A. Schuessler, \enquote{High-power mid-infrared frequency comb source
  based on a femtosecond er:fiber oscillator,} {\protect\JournalTitle{Opt.
  Lett.}} \textbf{38}, 2360--2362 (2013).

\bibitem{Vampa_OptLett2019}
G.~Vampa, S.~Vasilyev, H.~Liu, M.~Mirov, P.~H. Bucksbaum, and D.~A. Reis,
  \enquote{Characterization of high-harmonic emission from zno up to 11ev
  pumped with a cr:zns high-repetition-rate source,}
  {\protect\JournalTitle{Opt. Lett.}} \textbf{44}, 259--262 (2019).

\bibitem{Lee_OptLett2017}
K.~F. Lee, X.~Ding, T.~J. Hammond, M.~E. Fermann, G.~Vampa, and P.~B. Corkum,
  \enquote{Harmonic generation in solids with direct fiber laser pumping,}
  {\protect\JournalTitle{Opt. Lett.}} \textbf{42}, 1113--1116 (2017).

\bibitem{Ycas_OptLett2012}
G.~Ycas, S.~Osterman, and S.~A. Diddams, \enquote{Generation of a 660--2100 nm
  laser frequency comb based on an erbium fiber laser,}
  {\protect\JournalTitle{Opt. Lett.}} \textbf{37}, 2199--2201 (2012).

\bibitem{Li_RSI2016}
X.~Li, M.~A.~R. Reber, C.~Corder, Y.~Chen, P.~Zhao, and T.~K. Allison,
  \enquote{High-power ultrafast yb:fiber laser frequency combs using
  commercially available components and basic fiber tools,}
  {\protect\JournalTitle{Review of Scientific Instruments}} \textbf{87}, 093114
  (2016).

\bibitem{Arisholm_JOSAB2004}
G.~Arisholm, R.~Paschotta, and T.~S\"{u}dmeyer, \enquote{Limits to the power
  scalability of high-gain optical parametric amplifiers,}
  {\protect\JournalTitle{J. Opt. Soc. Am. B}} \textbf{21}, 578--590 (2004).

\bibitem{deOliviera_arXiv2019}
V.~S. de~Oliveira, A.~Ruehl, P.~Maslowski, and I.~Hartl, \enquote{Intensity
  noise optimization of a mid-infrared frequency comb difference frequency
  generation source,} {\protect\JournalTitle{arXiv:1904.02611}}  (2019).

\bibitem{Kowligy-inPrep}
A.~Kowligy, J.~Rutledge, A.~Catanese, M.~C. Silfies, S.~A. Diddams, and T.~K.
  Allison, \enquote{Efficient wideband uv/visible frequency comb generation via
  cascaded high-harmonic generation in lithium niobate waveguides,}
  {\protect\JournalTitle{In preparation}}  (2020).

\end{thebibliography}

\bibliographyfullrefs{main}

\end{document}